\documentclass{article}

\begin{document}
\title{Spatial Asymmetry For Particle Pairs And The Spin-Statistics Theorem}
\author{M. J. York\\975 S. Eliseo Dr. \#9, Greenbrae, CA 94904, USA\\{\it email}: mikeyork@jps.net}
\date{September 4, 1998, revision of October 5, 1998}
\maketitle
\begin{abstract}
We discuss the conditions under which identical particles may yet be distinguishable and the relationship between particle permutation and exchange. We show that we can always define permutation-symmetric state vectors. When the particles are completely indistinguishable, then exchange is equivalent to permutation and therefore the exchange eigenvalue for such permutation-symmetric state vectors is always $+1$. Exchange asymmetry arises when the particles are physically distinguishable, even though otherwise identical, and can be computed from the transformations that arise when the distinguishing features are reversed.

There is a fundamental spatial asymmetry between the relative orientations of any two vectors in a common frame of reference that persists even in the limit that the vectors coincide. For a pair of particles  this asymmetry between their spin quantization frames renders them distinguishable even when otherwise identical. In the conventional construction, this distinction is not properly accounted for. Particle exchange is then equivalent to reversing this relative orientation --- which requires a relative rotation by $2\pi$ on the spin quantization frame of one particle with respect to the other, thus resulting in the conventional exchange phase.

\end{abstract}


\section{Introduction}

There has recently been speculation that it is possible to prove the spin-statistics theorem without recourse to relativistic field theory. Duck and Sudershan \cite{Duck&Sud} have provided an extensive overview of such proofs and claim that a simple proof that introduces no new physical principle is not possible and go to great lengths to disprove various attempts that have been made, with the possible exception of a proposed proof by Berry and Robbins\cite{Berry&Rob}.

However, none of the discussion in \cite{Duck&Sud}, or papers cited there, seem to touch on the issues that the present author considers critical. (Although \cite{Berry&Rob} shows some similarities in the emphasis on single-valuedness and the derivation of the conventional exchange phase as originating in a $2\pi$ rotation, their claim of a single-valued wave function using 3-vectors rather than angular co-ordinates is not transparent to this author.) The issues critical for the present author are that:

\begin{enumerate}
\item Although identical particle symmetry is usually expressed in terms of the exchange phase, a unique exchange phase is only valid for unique (single-valued) state vectors. The experimentally verifiable aspect of the symmetry lies in the rules for quantum numbers of the allowed or excluded states (the exclusion rules). This is most easily summarized for fermions as the requirement that no two identical fermions may have the same quantum numbers. More general rules, for both bosons as well as fermions, can be expressed in terms of combined total spin\footnote{See, for example, Rose\cite{Rose}, chapter 12}. For instance, in a state in which all other quantum numbers are identical, the individual spins must combine such that the total spin quantum number $S$ must be even. 

\item Because the specification of quantum numbers is not sufficient for a single-valued state vector (in particular a rotation by $2\pi$ on the frame of reference of one particle may change the sign of a state vector but leave the quantum numbers unchanged) there is, in general, no simple unambiguous expression of the exclusion rules in terms of a unique exchange phase without the additional specification of accompanying information to eliminate the ambiguity.

\item The physical principle behind these exclusion rules lies in permutation invariance (nature does not care in which order we observe or describe the particles). Just as in classical physics, where permutation invariance reduces the number of allowed states of identical tossed coins for instance, this principle, together with properly accounting for any additional distinguishability which might arise from any physical asymmetry between the individual states, is the only physical principle we need to derive the exclusion rules.

\item The permutation operator has eigenvalues of $+1$ for order-independent, unique state vectors. Any other permutation phase has its origin in any order independence in the state vectors. Examples of such order-dependence would be arbitrary order-dependent phase factors (which we shall eschew) and permutation asymmetry in coupling coefficients when combining individual quantum numbers (which enables us to express the exclusion rules in terms of combined quantum numbers such as total angular momentum, for instance).

\item The exchange phase differs from the permutation phase only when there is an additional distinguishability which is not permuted with the other state variables (that is, where exchange is equivalent to {\it partial} permutation).

\item The identity of the particles and their indistinguishability are not the same. Although indistinguishable particles must be identical, the converse is not true. Just as two identical coins can be tossed in a way that distinguishes them, identical particle states can be described in ways that distinguish them. Furthermore, indistinguishability applies also to composite states of multiple particles, not just single particles, whether the constituent particles are identical or not and whether or not the total momentum lies on the mass shell of any known particle. Permutation invariance and the existence of an exchange phase applies to states of any two separately described particles, or composite systems, whether they are identical or not. The exclusion rules, however, arise only in the limit of indistinguishability.

\item There exists a fundamental asymmetry in two-particle states between the relative orientations of each particle (whether the orientation vector is the position vector, the momentum vector, the spin quantization axis or some other vector) which arises because the rotation which takes one particle's orientation vector into that of the other is physically distinguishable from its inverse. (If one rotation is clockwise then the other is counter-clockwise by the same amount.) {\it As a consequence it is not possible to choose a common frame of reference symmetrically between both particles}. This asymmetry was previously not recognized in the context of identical particle symmetry, to the author's knowledge, but we will describe it in detail in section \ref{sec:asymcommon}. It persists even in the case that both particles have their spin quantized in the same frame of reference or that any other vectors associated with the particles coincide. If not properly accounted for in the state variables (e.g. by using independent, symmetrically defined, frames of reference) this asymmetry distinguishes the individual particle states --- even if they are otherwise identical. In this case, uniqueness of the state vectors requires that exchange is a partial permutation that involves a reversal of the relative orientations. This is equivalent to a relative rotation by $2\pi$ on the frame of reference of one particle. This relative rotation is the origin of the conventional exchange phase. Of course, if the state variables did correctly account for the relative orientation, then the exchange phase would be merely the permutation phase: $+1$ (assuming independent quantum numbers for each particle). In all cases, whatever the exchange phase, the allowed states are permutation symmetric and the exclusion rules are the same. (The exclusion rules are, in essence, the physically observable consequence of the permutation symmetry and the relative orientation asymmetry.)
\end{enumerate}

In the next section we shall present a quick version of the exchange phase derivation, utilizing and clarifying these assertions. However, for the reader who feels that the relationship between permutation and exchange and the intrinsic asymmetry in two-particle states need further explanation, we provide this, in greater detail, in the subsequent sections.

\section{The Quick Version}

This section outlines the essential features of this derivation of the exchange phase in order to get the basic concepts over. The rest of the paper provides the details.

\subsection{Recapitulation}
The conventional argument about wave-function symmetry is that there exists an exchange operator $X$ such that if $\psi(\alpha,\beta)$ is the wave function for two identical particles ($a$ and $b$) with states described by the sets of variables $\alpha$ and $\beta$, then 
\begin{eqnarray}
\psi(\alpha,\beta) = X \psi(\beta,\alpha)
 = X^2 \psi(\alpha,\beta)
\end{eqnarray}
and hence the eigenvalues of $X$ are $\pm 1$.

However, the simple product wave function 
\begin{equation}
\psi(\alpha,\beta) = \psi(\alpha) \psi(\beta) \label{eqn:symmwavefunc}
\end{equation}
will always give an eigenvalue $+1$, whereas the experimentally observed exclusion rules lead us to believe (and we shall show that this is in fact dependent on certain implicit additional conventions - which we shall clarify) that for certain particles, the eigenvalue of $X$ is $-1$.

The normal explanation of this relies on conjecturing that the particles can be {\it labeled} ``1'' or ``2'' to {\it distinguish} them. Hence the product wave-function may be symmetric or anti-symmetric under interchange of these labels:
\begin{equation}
X \psi^1(\alpha) \psi^2(\beta) = \psi^1(\beta) \psi^2(\alpha)
= \pm \psi^1(\alpha) \psi^2(\beta)
\end{equation}

It is then stated, that without some extra physical principle, such as relativistic field theory, it is not possible to know which eigenstate of X applies. 

\subsection{A Contrarian View}
It is our view that the missing ingredient involves no new physical principle, but simply the recognition of the true significance of the labels ``1'' and ``2'' in a previously unnoticed physical asymmetry. When the state descriptions $\alpha$ and $\beta$ are not sufficiently unambiguous to make this asymmetry explicit (as is typically the case), then the labels serve this purpose instead.

If, on the other hand, $\alpha$ and $\beta$ are adequately unambiguous, then the labels can be dropped and we find we have a symmetric wave-function even for fermions - but the resulting exclusion rules are the same. (We use the terms boson and fermion to refer purely to integer or half-integer spin, not to any qualitative difference in the ``type'' of particle or the exchange phase of the wave function. Indeed they could apply equally to composite states of multiple particles, in which case they refer to the integer or half-integer value of the total angular momentum.)

Our contention is that the distinguishing labels ``1'' and ``2'' either have a physical significance in an asymmetry in the physical system (or, at least, in the way it is described) which therefore physically distinguishes the particles, {\it or they are irrelevant} and can therefore be dropped, leaving a symmetric wave-function, obeying eqn. \ref{eqn:symmwavefunc}, only, {\it even if the particles are identical fermions}. 

In other words, a non-trivial exchange operation only has meaning when the labels ``1'' and ``2'' have a physical significance that distinguishes the particles. When this is the case, the nature of the exchange operator and its eigenvalues {\it is determined by the physical relationship governing the distinguishing labels}.

A large part of the paper which follows is devoted to explaining and justifying this assertion of physical significance. For the purposes of this quick proof, the reader is invited to trust that this is so.

So the reader may now wish to ask: what aspect of the physical system enables us to distinguish particle ``1'' from particle ``2'' when the particles are presumed {\it identical}? And what bearing does this have on the exchange phase and the exclusion rules?

\subsection{Uniqueness Considerations}
Since the existence of a unique exchange phase depends on the construction of unique wave functions, let us look at the state variables involved in constructing a single-valued wave function. 

In co-ordinate space, the significant variables that concern us for a single particle are the position vector $\mathbf{r}$, the particle spin $s$ and its third component $m$. It is important to remember that the spin is quantized along the z-axis which is related to $\mathbf{r}$ by the orientation of $\hat{\mathbf{r}}$ in the frame of reference. In other words, the spin quantization frame is tied to the vector $\mathbf{r}$, by its orientation in that frame. 

However, the specification of these variables is well-known to be insufficient to define a single-valued wave function. If we rotate the frame of reference by $2\pi$, all these variables ($\mathbf{r},s,m$) are unchanged; yet because the spin-quantization frame of reference has been rotated, the wave function changes its phase by $(-1)^{2s}$. To obtain a single-valued wave function, we must specify the angular co-ordinates of $\hat{\mathbf{r}}$ over a wider range of polar angles than those limited to the physical space (or, equivalently the rotation which takes $\hat{\mathbf{r}}$ into the z-axis of the frame of reference). Thus if, instead of $\mathbf{r}$ we use $r$, $\theta$ and $\phi$, then we have a single-valued wave-function over the space $-\infty < \theta < \infty$ and $-\infty < \phi < \infty$, with the property that, for instance,
\begin{equation}
\psi(r,\theta,\phi+2\pi,s,m) = (-1)^{2s} \psi(r,\theta,\phi,s,m)
\end{equation}
when the rotation is about the z-axis. Note that if we had limited $\phi$ to the physical space ($0 \leq \phi < 2\pi$), then $\phi+2\pi$ would be equivalent to $\phi$ and we would not be able to distinguish the rotated wave function from the unrotated wave function. Hence, extending the range of $\theta,\phi$ is what enables us to obtain a single-valued wave function that distinguishes two frames related by a $2\pi$ rotation.

If, on the other hand, we had continued to use $\mathbf{r}$ without specifying the angular variables, then uniqueness would require us to specify a different wave function when the frame of reference is rotated by $2\pi$:
\begin{eqnarray}
\psi(\mathbf{r},s,m) & = & \psi(r,\theta,\phi,s,m)\\
\psi'(\mathbf{r},s,m) & = & \psi(r,\theta,\phi+2\pi,s,m) = (-1)^{2s} \psi(\mathbf{r},s,m)
\end{eqnarray}

Without distinguishing $\psi$ from $\psi'$, there is, therefore, a fundamental ambiguity in such wave functions. The same will be true for any other wave functions that depend only on variables that are unchanged by $2\pi$ rotations. Since however, it is not normally anticipated that particle exchange involves any such rotation on either particle, it is not usually thought necessary to eliminate this ambiguity when discussing exchange symmetry.

We take the opposite view --- which, as we shall show, is thoroughly justified by the results. To see this, one must be very careful to eliminate ambiguity. This involves the complete and unique specification of the state variables $\alpha$ and $\beta$ - including unambiguous specification of all angular orientations and the spin quantum numbers and quantization frame.

\subsection{The missing ingredient}
In a common frame of reference, there is an implicit asymmetry in the orientation of one particle with respect to the other. To see this, note that this relative orientation can be specified by a rotation by $\pm \pi$ about the axis which bisects their position vectors (in co-ordinate space) or which bisects their momentum vectors (in momentum space). For the rest of this section we shall work in co-ordinate space. If the rotation is chosen to be $+\pi$ for the orientation of $\mathbf{r}_b$ relative to $\mathbf{r}_a$ then it is $-\pi$ for the orientation of $\mathbf{r}_a$ relative to $\mathbf{r}_b$, since the latter must be the inverse of the former if the individual azimuthal angles in the plane perpendicular to the axis of rotation are to be preserved. Since each rotation is the inverse of the other, they cannot be the same and therefore are distinguishable. If one rotation is clockwise, the other must be counter-clockwise. (The uniqueness requirement is that the rotation $\mathbf{r}_a\rightarrow \mathbf{r}_b$ followed by $\mathbf{r}_b\rightarrow \mathbf{r}_a$ applied to either particle's wave function should recover the original wave function. Hence $\mathbf{r}_b\rightarrow \mathbf{r}_a$ is the inverse of $\mathbf{r}_a\rightarrow \mathbf{r}_b$. If we had chosen $+\pi$ for both rotations then $\mathbf{r}_a\rightarrow \mathbf{r}_b$ followed by $\mathbf{r}_b\rightarrow \mathbf{r}_a$ would result in a $2\pi$ rotation on $\mathbf{r}_a$ and potentially change the sign of the wave function of particle $a$.)

We shall go into more detail about this asymmetry in section \ref{sec:spatasym}.

For the present discussion, all we need to know is that a single-valued two-particle wave function requires the unique specification of the relative orientation of one particle to the other in a common frame of reference --- otherwise a $2\pi$ rotation on one particle's vector relative to the other is equivalent to a rotation by $-2\pi$ on that particle's frame of reference --- in particular, the spin quantization frame --- with an accompanying phase change, but not on the other's.

Clearly this unique specification can be done using the unambiguous variables ($r,\theta,\phi,s,m$) for each particle since the angular co-ordinates fix the relative orientation of $\mathbf{r}_a$ and $\mathbf{r}_b$. If, instead we were to use the ambiguous variables ($\mathbf{r},s,m$) we would need some other method, such as additional labels ``1'' and ``2'' to fix the relative orientation of ``1'' to ``2''. In the latter case, as we shall show, the relative orientation of $\mathbf{r}_a$ to $\mathbf{r}_b$ is reversed by the exchange --- which is equivalent to rotating one particle's frame of reference by $\pm 2\pi$. 

Omitting (or ignoring) this relative orientation means that the wave function will not be unique and the exchange phase indeterminate.

In case the reader is feeling a little confused, note that, if the relative orientations of the particles are defined both before and after exchange by their labels ``1'' and ``2'', then they cannot also be defined by their angular co-ordinates, and {\it vice versa}.

To see this, consider a frame of reference in which the z-axis bisects the position vectors associated with the individual particles. If, instead of fixing $\mathbf{r}_b\rightarrow \mathbf{r}_a$, we fix $\mathbf{r}_2\rightarrow \mathbf{r}_1$, then the azimuthal angles (with respect to the axis of rotation) of the two particles are related by
\begin{equation}
\phi^2 = \phi^1 + \pi
\end{equation}
Then when particle ``1'' is $a$,
\begin{equation}
\phi^2_b = \phi^1_a + \pi
\end{equation}
whereas, when particle ``1'' is $b$,
\begin{equation}
\phi^2_a = \phi^1_b+ \pi
\end{equation}
Hence, we cannot simultaneously choose both 
\begin{equation}
\phi^2_a = \phi^1_a
\end{equation}
and
\begin{equation}
\phi^2_b = \phi^1_b
\end{equation}
and thus, if we exchange the labels yet preserve their significance, then the state variables of at least one of the particles must change.

Specifically, if we define
\begin{equation}
\Delta_{ab} = \phi_b - \phi_a
\end{equation}
then we see that $\Delta_{ab}$ will change from $+\pi$ when $a=1$ to $-\pi$ when $b=1$. For example, if we leave the state variable $\phi_a$ unchanged by the exchange, then we find that $\phi_b$ must change to $\phi_b' = \phi_b-2\pi$ in changing from particle ``2'' to particle ``1''. Alternatively, if we leave the state variable $\phi_b$ unchanged by the exchange, then we find that $\phi_a$ must change to $\phi_a' = \phi_a+2\pi$ in changing from particle ``1'' to particle ``2''.

To recover the original state variables, and maintain the same common frame of reference for both particles, we must therefore rotate the frame of reference of one particle by $\pm 2\pi$ about the bisecting axis.

The implicit distinction between ``1'' and ``2'' that we have determined is equivalent to:
\begin{equation}
\psi^1(\mathbf{r}_a,s_a,m_a) \psi^2(\mathbf{r}_b,s_b,m_b) = \psi(r_a,\theta_a,\phi_a,s_a,m_a) \psi(r_b,\theta_b,\phi_a+\pi,s_b,m_b)
\end{equation}
where the condition $\phi_b = \phi_a + \pi$ is necessary for the unique specification of the relative orientations. Applying the exchange operator in such a way as to maintain $\phi_b$ unchanged, we find
\begin{eqnarray}\label{eqn:quickexchange}
X \psi^1(\mathbf{r}_a,s_a,m_a) \psi^2(\mathbf{r}_b,s_b,m_b) = \psi^1(\mathbf{r}_b,s_b,m_b) \psi^2(\mathbf{r}_a,s_a,m_a)\nonumber\\
 = \psi(r_b,\theta_b,\phi_a+\pi,s_b,m_b) \psi(r_a,\theta_a,\phi_a+2\pi,s_a,m_a)\nonumber\\
 = (-1)^{2s_a}\ \psi^1(\mathbf{r}_a,s_a,m_a) \psi^2(\mathbf{r}_b,s_b,m_b)
\end{eqnarray}
Alternatively, the interchange $1\leftrightarrow2$, keeping $\phi_a$ fixed is equivalent to replacing $\phi_b$ by $\phi_b-2\pi$. Either way, for identical particles ($s_a=s_b=s$), the eigenvalue of $X$ for these order-dependent wave-functions is $(-1)^{2s}$. Whereas the eigenvalue of $X$ for the wave function $\psi(r_a,\theta_a,\phi_a,s_a,m_a) \psi(r_b,\theta_b,\phi_b,s_b,m_b)$, where both $\phi_a$ and $\phi_b$ (and the relationship between them) are unchanged by $X$, is always $+1$.

The reader is invited to read the rest of this paper for a more detailed explanation and to satisfy themselves that no smoke or mirrors are involved.

\section{Permutation Invariance And Order-Free Notation}	
\label{sec:PI}

This section will summarize the notion of permutation invariance for multi-particle states whether identical or not. This may puzzle the reader since the connection between particle permutation and identical particle exchange is not often addressed. However, it should be apparent, or we hope will soon become so, that when the particles are not only identical but also fully {\it indistinguishable} then particle exchange is equivalent to permutation. That is, exchange of indistinguishable particles is just a special case of permutation. Hence by studying the general case of particle permutation, whether distinguishable or not, and whether identical or not, and the physical features that may distinguish identical particles, we shall obtain some rules that enable us to understand particle exchange for identical particles, whether they are distinguishable or not, and to discover the symmetry properties therein.

Although this connection and these distinctions are not usually expressed in this way, by doing so, and using an appropriately general and unambiguous notation, we hope not to fall into the subtle traps that await the unwary. Although this may seem unnecessarily pedantic in places, the reader is begged to bear with the process. 

Conventional discussion of identical particle exchange in quantum mechanics, because of its complex historical legacy, can very easily lead to the widely-held belief that the exchange symmetry of state vectors for identical particles cannot be further determined without recourse to additional assumptions about nature, such as relativistic quantum field theory. This is in spite of the fact that no such additional assumption is necessary in classical physics (e.g. where permutation invariance {\it by itself} reduces the number of possible independent states when tossing two identical coins from four permutations when the coins are distinguishable to three combinations when the coins are {\it in}distinguishable). It is our contention that, in quantum mechanics, just as in classical mechanics, the reduction in the number of states of two identical particles can be computed simply by recognizing any distinguishability in the system and properly accounting for it.

\subsection{Permutation Invariance}
We define our permutation invariance assumption as:

\newtheorem{axiom}{Axiom}
\begin{axiom}[Permutation Invariance]
The physical properties of multiple entity states are independent of the order in which we observe or describe the collection of individual entities that make up the whole state.
\end{axiom}

The reader may consider this to be obvious and hardly worthy of explicit statement. However, we have chosen to make it explicit because some descriptions of exchange asymmetry in quantum mechanics (e.g. those which insist that the exchange phase is always $-1$ for fermions and not a matter of convention) actually violate this principle, as we shall show.

This permutation assumption applies equally to classical physics as to quantum physics. It also applies equally to distinguishable entities as to indistinguishable entities. It can be expressed mathematically in terms of individual state descriptions and collections of such state descriptions that do not necessarily have any relation to quantum mechanical state vectors.

Suppose we have several possibly distinguishable entities labeled $i,j,k...$. These could be individual particles, or more complex entities. They could be distinguishable either by type (identity) or method of description. Suppose that their physical states are described by $S^i_a,S^j_b,S^k_c...$. These descriptions state that the entity distinguished by label $i$ is in a state $S_a$ and the entity distinguished by label $j$ is in a state $S_b$ and so on. We do not, at this stage, nor the whole of this section and section \ref{sec:PA}, have to know anything about how these entities (or their states) are actually described (their state variables,etc), just that they {\it can} be described and that they can be distinguished if their labels are different and, for section \ref{sec:PA}, that they have corresponding quantum mechanical state vectors in Hilbert space. Suppose also that these individual state descriptions are independent of each other and of the order in which we describe the entity states. The combined state is then described by a collection of individual states, which we can write as a list:
\begin{equation}
S^{ijk...}_{abc...} = S^i_a;S^j_b;S^k_c;... \label{eqn:collectdef}
\end{equation}

It is in the nature of lists that they are {\it ordered}. Although this collection is written as such a list, our permutation invariance assumption is that the properties of the collection are independent of the listing order. Hence all lists related by permuting the order in which the individual states appear are equivalent:
\begin{equation}
S_{abc...}^{ijk...} \equiv S_{bac...}^{jik...} \equiv etc \label{eqn:perminv}
\end{equation}
and any one such list can stand in for any other as a description of the complete state. In other words, the number of independent collections is given by the number of {\it combinations} of entity states rather than the number of permutations.

In the case that two entities are indistinguishable, we simply equate their labels ($i=j$). Then we find the additional property that
\begin{equation}
S_{abc...}^{iik...} \equiv S_{bac...}^{iik...} \equiv etc \label{eqn:redundantsuper}
\end{equation}

(Note that, in this case, permutation is equivalent to ``exchanging'' the state descriptions of the indistinguishable entities. We shall go into this in more detail in section \ref{sec:PA}.)

An important consequence of this indistinguishable entity symmetry (eqn. \ref{eqn:redundantsuper}) is a reduction in the number of independent collections. This is well-known in the case of coin-tossing to reduce the number of distinguishable combinations of two indistinguishable coins from four to three. In quantum mechanics it is known that this symmetry is connected to the identical particle exclusion rules but it is widely believed that these rules cannot be found from permutation invariance alone. Our purpose is to show that these exclusion rules can indeed be determined purely from permutation invariance as in the classical case --- as long as all physical distinguishability is properly accounted for.

\subsection{Order-Free Notation}
It should be pointed out, however, that even with distinguishable entities, eqn. \ref {eqn:perminv} implies that we still have a redundancy in our notation. If we had a notation that had no order-dependence in it, we could remove this redundancy. 

We could, for instance, write the individual states over the top of each other to illustrate the absence of any significance to the order in which we describe the individual entities. However, with normal two-dimensional paper or computer screen layouts this is likely to result in illegibility.

In the case where all entities are distinguishable, we could arbitrarily choose to always use a particular ordering for the entities to remove the redundancy. For example, we could choose to always list the labels in a particular order, whether entities with those labels are present or not. However this will not work when we have indistinguishable entities, because we will still end up with redundant state descriptions obeying eqn. \ref{eqn:redundantsuper}.

One notation that could remove this redundancy for indistinguishable entities would be to use a table for each type of distinguishable entity, labeled by some arbitrary ordering of the allowed states for that entity, and in which the table entries simply specified the number of entities present in the state identified by the index (a single entity state description). For example, a state of two indistinguishable tossed coins could be represented by a table with two columns labeled $H$ and $T$ (one for each allowed state). The entry in each column would be the number of coins in that state. Hence, instead of listing four states in the ordered notation: $H;H$, $H;T$, $T;H$ and $T;T$, we would find that there are only three collections describable in the unordered notation: $2;0$, $1;1$, and $0;2$.

The reason for drawing the readers attention to this alternative order-free notation is to point out that in principle it is just as possible to use such a notation to describe quantum-mechanical states as it is to describe classical states such as tossed coins. This gives us a method of constructing order-free state vectors and therefore leads to the important conclusion that {\it any order-dependence in quantum-mechanical state vectors is an a consequence of the method of constructing the state vectors and the notation used and has no other significance}. In the case that the particles are indistinguishable ($i=j$), permutation is equivalent to exchange and, if the state vectors are constructed in an order-free way, then the exchange eigenvalue will be $+1$. Hence, contrary to popular misconception, there is nothing sacred about the exchange phase of $-1$ and it arises simply because of an implicit order-dependence in the conventional construction.

Clearly this would be a cumbersome notation when the number of allowed entity states is much greater than the number of entities and requires an efficient method of abbreviation when describing entities with continuous quantum numbers, but the point is still the same: order-dependence of quantum-mechanical state vectors is a consequence of the choices made in constructing the state vector --- whether dealing with distinguishable particles or indistinguishable particles. The rest of this paper will show how such order-dependence arises in conventional ways of describing particle states and the notations used.

\section{Phase Ambiguity In State Vectors}
\label{sec:PA}

A common source of confusion when discussing the spin-statistics theorem is the notion that there is an arbitrary phase multiplier for any state vector. This often leads to the supposition that the exchange phase relating two identical particle state vectors which differ only in the ordering of the individual particles cannot be uniquely determined without some additional assumption.

The argument goes that particle exchange is a new discrete operation $X$ which can change the phase of the state vector. Since a repeat exchange recovers the original state vector ($X$ is its own inverse), then the eigenvalues of $X$ are $\pm 1$.

Our claim here, however, is that unless exchange of indistinguishable particles has some new unknown physical significance, then it is nothing more than permutation of the individual particle descriptions in the state vector which {\it by itself} can be of no significance in a state description. 

To see this, let us define a permutation operator $P$ such that:
\begin{equation}
P\ |S_{ab}^{ij}> = |S_{ba}^{ji}>
\end{equation}
whereas the exchange operator $X$ is such that:
\begin{equation}
X\ |S_{ab}^{ij}> = |S_{ba}^{ij}>
\end{equation}

Clearly, in the case $i = j$, we find $X$ and $P$ have the identical effect.

We have also shown that it is possible to define order-free state descriptions. Since these can also have quantum mechanical state vectors assigned to them, it is clearly possible to define order-free state vectors for which permutation is therefore the {\it identity} operation and hence that the eigenvalue of $P$ is always $+1$ and, therefore, in the case of indistinguishability, the eigenvalue of $X$ is also $+1$. Hence the listing order can be relevant only when it is linked to some order-dependence in the individual particle state descriptions and/or their notation and particle exchange $X$ can have an eigenvalue of $-1$ only when the particles are distinguishable. Some further explanation or qualification is clearly necessary. In particular we need to examine the relationship between identity and indistinguishability, the {\it uniqueness} properties of state vectors and look closely at the significance of phase ambiguity.

\subsection{Identity And Indistinguishability}\label{sec:IdandInd}
We need to point out that, as we have used the term in the previous subsection (and throughout this paper), indistinguishability is not the same thing as identity.
Usually, in quantum mechanics, we define any two particles to be {\it identical} if they are permitted the same range of physical states - i.e. the same range of quantum numbers, continuous or discrete (e.g. momentum on a unique mass shell, unique charge, etc.). Identical particle states are, of course, distinguished by the actual values of the quantum numbers (e.g. linear or angular momentum). However, even in the case of identical states, identical particles may still be distinguishable if their physical states are differently described (such as, for instance, a difference in the method of choosing their frame of reference) and, in what follows, we shall assume that this distinction is contained in the distinguishing labels ($i,j$) and any particle identity, if relevant, is implicit in its quantum numbers. 

Throughout, the rest of this paper, the reader should be aware, therefore, that it is particle indistinguishability (which implies $i=j$ as well as identical quantum numbers) rather than identity, which is significant in determining the exclusion rules. Particle identity, of itself, is of no significance except that if two particles are indistinguishable then they must also be identical, although the converse is not true because identical particles may still be distinguishable by their variable quantum numbers or the way their states are described.

By the same token, none of what follows is specific to states of individual particles. Everything in this paper is equally true for composite states (and permutations of pairs of composite states) which have no specific mass shell or ``identity''. In determining the permutation or exchange phase, and the physical consequences of permutation invariance, we concern ourselves only with the state variables for each entity and any additional distinguishability implied by the labels $i,j$. Hence the exclusion rules apply also to pairs of such composite states and the identities of any individual constituent particles are irrelevant, except where they are a part of the state descriptions. (Composite states may be states of arbitrary constituents or specific constituents --- it doesn't matter which.) 

From now on, therefore, our discussion will centre around the state variables and distinguishability necessary to uniquely describe the state. Any mention of particle identity is incidental and of no critical importance.

\subsection {Uniqueness And Equivalence Of State Vectors}\label{sec:unique}
Although it is true we can choose our state vector for any given state from an infinite set of state vectors that differ only by a phase factor, we are always free to choose {\it one} such vector to be the unique representative of our physical state. (Not only that, but we {\it must} make such a choice if we wish to calculate effects such as interference.) Once we have a unique prescription for making that choice for any state, then we no longer have any phase ambiguity unless we change the way we describe the state. Hence any residual phase ambiguity arises solely from any ambiguity in the state description. The proof of this is trivial. Suppose $S$ and $S'$ are two alternative descriptions of the same physical state. $S$ can be a single particle state or a list of such states. Then
\begin{equation}
|S'> = f(S',S)\ |S>
\end{equation}
where $f(S',S)$ is a phase factor. Clearly, uniqueness requires 
\begin{equation}
f(S,S) = 1. \label{eqn:uniqueness}
\end{equation}
So the existence of a phase change depends on the distinction between the state {\it descriptions} $S'$ and $S$, even though they represent the same physical state.

A well-known example of such a phase change occurs in the angular momentum representation (i.e. when we have states of definite $j$ and $m$) when we rotate the frame of reference about the angular momentum quantization axis. Although the frame of reference (which is part of the state description) changes ($S\rightarrow S'$), the physical quantum numbers remain unchanged. However, if both  $|S>$ and $|S'>$ lie in the same Hilbert space,  yet the transformation is of the observer only (and leaves the physical state unchanged) they may differ at most by a phase. In fact that phase is uniquely determined by the angle of rotation and the third component of angular momentum.

Now suppose we have two descriptions of the same state that are physically {\it equivalent} ($\bar S \equiv S$). By this we mean that not only are the quantum numbers identical, but so are all other physical features of the state, such as the frame of reference. Since there is no change in frame of reference or any other physical transformation and no other physically observable difference, there can be no transformation in Hilbert space to correspond to the change in description $S\rightarrow \bar S$. If both state vectors exist in the same Hilbert space and yet cannot be related by any transformation except the identity operation then they must be identical:
\begin{equation}
|\bar S> = |S> \label{eqn:equiv}
\end{equation}

As an alternative proof, suppose that there {\it was} a phase difference ($f(\bar S,S)\not= 1$). Then the relationship between $\bar S$ and $S$ would be as physically significant as any physical transformation that produced the same change in phase - which would violate our definition of $\bar S$ that there was no such physical significance. (As an example of this, consider a change in description which produces $f(\bar S,S)=e^{i\eta}$ for a state of definite angular momentum. This phase change is also produced by a rotation by $-\eta/m$ where $m$ is the component of angular momentum along the axis of rotation. Hence the change of description $S\rightarrow \bar S$ is equivalent to such a rotation. Hence our transformation $S\rightarrow \bar S$ must be physically significant. This is true for a single-particle state or a multi-particle state and, in the latter case, whether or not the particles are identical.)

The consequence of this is that we can always choose our state vectors so that any residual phase factor between two state vectors for the same state can be limited to situations where there is a physical difference between the ways we observe and describe the states (i.e. where this difference can be described by a transformation in Hilbert space) and not from a change in notation only. Indeed, uniqueness of the state vector for a given state description requires that we {\it must} choose our state vectors in this way.

In summary, if state vectors are to be unique, then state vectors for state descriptions that are physically equivalent must be identical. In other words, indistinguishable states have identical state vectors. States that are related by a physical transformation, however, have state vectors that are related by an equivalent transformation in Hilbert space. When the physical transformation leaves the quantum numbers unchanged, then it describes the relationship between the additional features which distinguish the state descriptions, such as the labels introduced in the previous section, and the Hilbert transformation defines the phase change that relates the different state vectors. (In the case of a rotation by $2\pi$, note that it is still a physically recognizable transformation, even if the resulting physical state is not recognizably different, except in the context of how we got to it.)

In particular, unless you consider particle permutation (or, in the case of indistinguishability, ``particle exchange'') to be a physically significant transformation, (contrary to our permutation invariance assumption) then uniqueness implies that it cannot, {\it by itself}, introduce any permutation phase (or ``exchange phase'') between multi-particle state vectors unless we introduce such a phase, whether explicitly or implicitly, by a physically significant order-dependent asymmetry in the individual particle state descriptions.

\subsection{The Origin Of An Exchange Phase}\label{sec:orgnxphase}
If our listing-order is purely a matter of notation and has no significance for the description of the individual states, then we have seen that uniqueness requires that our state vectors are order-independent and, in the case of indistinguishable particles, this permutation symmetry alone will be a filter for the permitted states. We shall now show how a permutation (or exchange) phase can nevertheless arise in situations where we use a notation in which the particle ordering affects the individual descriptions of the individual particles. Then the exchange phase of the state vectors will be determined by the Hilbert space transformations brought about by {\it changes in the individual descriptions} resulting from the change in ordering.

In general, using the ordered notation, it is apparent that eqns. \ref{eqn:perminv} and \ref{eqn:equiv} (permutation invariance and uniqueness of state vectors for equivalent state descriptions) imply that

\begin{equation}
|S^{ij...}_{ab...}> = |S^i_a;S^j_b;...> =  |S^j_b;S^i_a;...> = |S^{ji...}_{ba...}> \label{eqn:orderindep}
\end{equation}

Now, we saw in subsection \ref{sec:unique} that state vectors for distinguishable particles are related by the transformations in Hilbert space corresponding to the physical transformations that relate their distinguishing state descriptions. Hence, even for states with identical quantum numbers, such as identical particles in the same physical state, this distinguishability implies that
\begin{eqnarray}
S_a^i \not\equiv S_a^j\\
S_b^i \not\equiv S_b^j
\end{eqnarray}

For such states, where the distinguishability results purely from a difference in description for what is ostensibly the same physical state, then we can define both state vectors in the same Hilbert space and differing by, at most, a phase:
\begin{eqnarray}
|S_a^i> & = & f(S_a^i,S_a^j)\ |S_a^j>\nonumber\\
|S_b^i> & = & f(S_b^i,S_b^j)\ |S_b^j>\label{eqn:singlephase}\\
|S_{ab...}^{ij...}> & = & f(S_{ab...}^{ij...},S_{ba...}^{ij...})\ |S_{ba...}^{ij...}>\nonumber\\
|S_{ab...}^{ij...}> & = & f(S_{ab...}^{ij...},S_{ab...}^{ji...})\ |S_{ab...}^{ji...}> \label{eqn:doublephase}
\end{eqnarray}
and we find that
\begin{eqnarray}
f(S_{ab...}^{ij...},S_{ba...}^{ij...}) & = &  f(S_{ab...}^{ij...},S_{ab...}^{ji...})
\end{eqnarray}
whereas, from eqn. \ref{eqn:orderindep}
\begin{eqnarray}
f(S_{ab...}^{ij...},S_{ba...}^{ji...}) & = & 1
\end{eqnarray}
{\it In other words, any exchange phase that might arise is a consequence not of the particle permutation but of the exchange of distinguishing characteristics ($i\leftrightarrow j$).}

In the next section we address the question of the relation between the single-particle distinguishability phase factors and the two-particle exchange phase factors.

\subsection{Determination Of The Exchange Phase}\label{sec:detxphase}
To relate the multi-particle exchange phases to single-particle distinguishability phases we need to know how to relate transformations in multi-particle space to those in single-particle space. We do this by relating the multi-particle state vectors to the single-particle state vectors. 

Multi-particle state vectors can be chosen as direct product state vectors from the Hilbert space that arises from the direct product of the single-particle Hilbert spaces, $H^i$, $H^j$ etc. For distinguishable particles ($i\ne j$), always described in a particular order ($i$ first), the state vector will lie in the direct product space $H^i\otimes H^j$. If described in reverse order ($j$ first) then it would lie in $H^j\otimes H^i$. These two different product vectors lie in two distinct spaces. To define a permutation invariant and order-independent state vector obeying eqn. \ref{eqn:orderindep} we take the symmetrized linear combination of the direct product vectors:
\begin{equation}
|S_{ab}^{ij}> = \alpha (|S_a^i>|S_b^j>\ +\ |S_b^j>|S_a^i>) \label{eqn:directprod1}
\end{equation}
where the factor $\alpha$ is for normalization only. For two particles, distinguishable by the labels $i,j$, we also have
\begin{equation}
|S_{ba}^{ij}> = \alpha (|S_b^i>|S_a^j>\ +\ |S_a^j>|S_b^i>) \label{eqn:directprod2}
\end{equation}
and substituting the single particle distinguishing phases of eqn. \ref{eqn:singlephase} in eqn. \ref{eqn:directprod1} we find:
\begin{equation}
|S_{ab}^{ij}> = \alpha (|S_a^i>|S_b^j>\ +\ f(S_b^j,S_b^i)\ f(S_a^i,S_a^j)\ |S_b^i>|S_a^j>) 
\end{equation}
which enables us to compute the two-particle exchange phase in eqn. \ref{eqn:doublephase}:
\begin{equation}
f(S_{ab...}^{ij...},S_{ba...}^{ij...}) = f(S_a^i,S_a^j)\ f(S_b^j,S_b^i) = {f(S_a^i,S_a^j) \over f(S_b^i,S_b^j)}
\end{equation}

{\it Thus the exchange phase is computed purely from the single particle phase changes that arise from the exchange of distinguishing features $i$ and $j$, and we have indicated how it can be done in the general case without introducing any special additional assumptions such as relativity or local field theory.}

However, we would stress that the only reason for introducing these exchange phase factors is when there is a genuine physical asymmetry (which implies distinguishability) in the way the individual particle states are described, corresponding to a transformation in Hilbert space when the distinguishing features are exchanged. Without exchanging such distinguishing features of the individual state descriptions, uniqueness requires that the phase obtained by simple re-ordering would always be unity.

We shall now turn our attention to how the conventional exchange phase arises and the implicit conventions in the choice of state vectors that render the particles distinguishable. An implicit order dependence in the conventional description of quantum-mechanical two-particle states, which arises from an asymmetry in the distinguishing spatial rotations necessary when we choose a common frame of reference, and necessitates a transformation in Hilbert space when we re-order the particles, is the origin of the asymmetry conventionally expressed in the spin-statistics theorem. State vectors that were defined symmetrically (so that identical particles, for instance, were also indistinguishable, except by their quantum numbers) would still be symmetrical under ``particle exchange'' and still produce the same exclusion rules.

\section{Spatial Asymmetry And Permutation Invariance}\label{sec:spatasym}
We have previously and frequently alluded to an inherent spatial asymmetry in two-particle states. In this section we shall provide a detailed discussion of this asymmetry and its consequences for defining unique state vectors.

\subsection{Asymmetry In A Common Frame Of Reference}\label{sec:asymcommon}
My purpose here is to explain why it is not possible to choose a common frame of reference for two particles in a way that is symmetrical (does not distinguish) between the orientations of the individual particles. One can choose either a common frame of reference or distinct symmetrically defined frames of reference for each particle. But one cannot do both simultaneously. If this asymmetry (distinction) in a common frame is not properly accounted for in the state variables, then it must be specified in some other way (e.g. as a distinguishing label --- implying a corresponding exchange asymmetry) if we are to have uniquely defined state vectors.

The state description of each particle in the system has at least one physical vector $\mathbf{v}$ attached to it. It could be the position vector $\mathbf {r}$ (in co-ordinate space), the linear momentum vector $\mathbf {p}$ (in linear momentum space), the spin quantization axis or any other physical vector that is part of the state description of that particle. In a common frame of reference, each vector is described with respect to that common frame.

In quantum mechanics, a unique state vector for a single particle requires the unique specification of the rotation $R(\hat{\mathbf{z}}\rightarrow \hat{\mathbf{v}})$ which takes the z-axis of the frame of reference into its physical vector $\mathbf{v}$ or {\it vice versa}. For instance, in momentum space, the state vector for arbitrary momentum is defined by taking the rest frame state vector, applying a boost along the z-axis and then a rotation which takes this z-axis into the direction of motion. (Conventionally, the boost is expected to be a Lorentz boost. For our purposes this does not matter --- a Galilean boost would be just as valid. Nothing in this paper depends on Lorentz invariance.) Similarly, in choosing a spin quantization frame, we must uniquely specify the rotation which takes the z-axis into the spin quantization axis.

In a two-particle state, this has to be done for {\it both} particles. For unique, individual state vectors these rotations ($R_a=R(\hat{\mathbf{z}}\rightarrow \hat{\mathbf{v}}_a)$ and $R_b=R(\hat{\mathbf{z}}\rightarrow \hat{\mathbf{v}}_b)$) must be defined independently for each particle or the difference must be accounted for in the state descriptions. The question then arises: Is it possible to define these orientations in a way that is symmetric between both vectors and obtain the same frame of reference for both particles? And, if not, what are the consequences for the state vector in a common frame of reference?

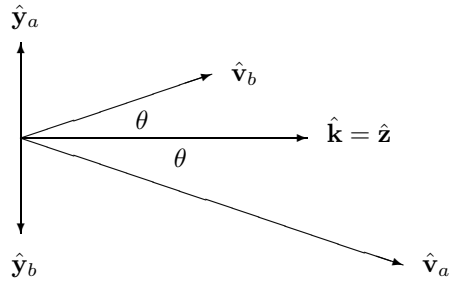
\begin{figure}[htbp]\begin{center}
\caption{\label{fig:vvx} Symmetric z-axis}
\setlength{\unitlength}{1in}
\begin{picture}(4,1.8)
\put(1,1){\vector(0,1){0.5}}
\put(0.95,1.6){$\hat{\mathbf{y}}_a$}
\put(1,1){\vector(3,1){1.0}}
\put(2.1,1.3){$\hat{\mathbf{v}}_b$}
\put(1.6,1.05){$\theta$}
\put(1,1){\vector(3,0){1.5}}
\put(2.6,0.98){$\hat{\mathbf{k}}=\hat{\mathbf{z}}$}
\put(1.8,0.85){$\theta$}
\put(1,1){\vector(3,-1){2.0}}
\put(3.1,0.3){$\hat{\mathbf{v}}_a$}
\put(1,1){\vector(0,-1){0.5}}
\put(0.95,0.3){$\hat{\mathbf{y}}_b$}
\end{picture}
\end{center}
\end{figure}

Let us start by choosing a symmetrically defined z-axis. This is easy to do. We choose the axis $\mathbf{k}$ which bisects the two vectors. Each vector will then make an angle
\begin{eqnarray}
\theta = {\cos^{-1}(\hat{\mathbf{v}}_a.\hat{\mathbf{v}}_b)\over 2}
\end{eqnarray}
with the common z-axis. (See fig. \ref{fig:vvx}.) It matters not whether we choose $0 \le \theta < {\pi\over 2}$ or $-\pi \le \theta < -{\pi \over 2}$ {\it as long as we preserve symmetry by making the same choice for both particles}. 

Now let us see if we can choose a common y-axis symmetrically. Like the z-axis, this must lie in the plane which bisects $\hat{\mathbf {v}}_a$ and $\hat{\mathbf {v}}_b$. It must also be perpendicular to the z-axis. Hence it must be given by either
\begin{equation}
\hat{\mathbf{y}} = \hat{\mathbf{v}}_a\times\hat{\mathbf{v}}_b
\end{equation}
or
\begin{equation}
\hat{\mathbf{y}} = \hat{\mathbf{v}}_b\times\hat{\mathbf{v}}_a
\end{equation}

Each choice of y-axis and x-axis is asymmetric between the particles. If we choose
\begin{equation}
\hat{\mathbf{y}}_a = \hat{\mathbf{v}}_a\times\hat{\mathbf{v}}_b
\end{equation}
then
\begin{equation}\label{eqn:symydef}
\hat{\mathbf{y}}_b = \hat{\mathbf{v}}_b\times\hat{\mathbf{v}}_a = -\hat{\mathbf{y}}_a
\end{equation}
and we can independently and symmetrically define $\hat{\mathbf{y}}_a$ and $\hat{\mathbf{y}}_b$, but, since they are opposite to each other, they cannot coincide. 
Furthermore, the x-axis can then not be in the bisecting plane, and therefore favors one particle over the other.

This asymmetry ($\hat{\mathbf{y}}_b = -\hat{\mathbf{y}}_a$) persists even in the limit $\hat{\mathbf {v}}_a\rightarrow \hat{\mathbf {v}}_b$. Even with a common z-axis, we cannot also choose a common y-axis without preferring one particle over the other. Hence {\it we cannot choose a common frame of reference for both particles without introducing an asymmetry (and therefore a possible exchange phase) between the particles}.

\subsection{Accounting For The Asymmetry}\label{sec:asymaccount}
The existence of this asymmetry means that we cannot, simply by specifying a common frame, assume that state vectors for two particles are both unique and order-independent, since the method of specifying the orientation of each particle in that common frame will differ and therefore imply an asymmetry that, if not accounted for, will violate uniqueness. However, it is possible to account for this asymmetry and examine its effect on our state vectors, if we use the following prescription:

\begin{enumerate}
\item Define independent (but symmetrically chosen) frames of reference for each particle. We know then that the frames of reference do not introduce any new asymmetry. Hence the individual state vectors for each particle are uniquely defined and the two-particle state vector can then be defined by the symmetrized direct product to be permutation-symmetric {\i and} exchange symmetric --- as long as every other detail of the single-particle state vectors has also been defined symmetrically. Furthermore, as long as there are no distinguishing features not explicit in the state variables, these state vectors will also be exchange symmetric. 
\item By examining the effect of the rotations which then take each particle's independent frame of reference into a common frame we can compute the effect of the asymmetry in the common frame.
\end{enumerate}

We shall now turn our attention to defining independent frames of reference in a way that is symmetrical between the particles. {\it By defining the individual state vectors in their respective independent frames, we know that the individual state vectors can be uniquely defined in a way that does not discriminate (distinguish) between the particles and, therefore, we can define a permutation-symmetric and exchange-symmetric two-particle state vector.}

\subsection{Independent ``Parallel'' Frames}\label{sec:parallel}
For instance, we can choose the z-axis for each particle to be parallel to its vector $\hat{\mathbf{v}}$:
\begin{eqnarray}\label{eqn:parallelz}
\hat{\mathbf{z}}_c = \hat{\mathbf{v}}_c
\end{eqnarray}
and the y-axes by their cross-product:
\begin{equation}\label{eqn:currenty}
\hat{\mathbf{y}}_c = \hat{\mathbf{v}}_c\times\hat{\mathbf{v}}_o
\end{equation}
where $c$ is the {\it current} particle and $o$ is the {\it other} particle. Clearly, this gives us a symmetric method for choosing independent frames. We shall call this choice of y- and z-axes, for each particle, its {\it parallel} frame. 

\subsection{Independent ``Bisecting'' Frames}\label{sec:bisecting}
Alternatively, choosing a common z-axis as the axis $\hat{\mathbf{k}}$ which bisects the two vectors:
\begin{eqnarray}\label{eqn:bisectingz}
\hat{\mathbf{z}}_a = \hat{\mathbf{z}}_b = \hat{\mathbf{k}}
\end{eqnarray}
and the y-axes again by their cross-product (eqn. \ref{eqn:currenty}) gives us another symmetric method for choosing independent frames. We shall call this choice of y- and z-axes, for each particle, its {\it bisecting} frame. 

\subsection {General Symmetrically-Defined Independent Frames}
The parallel and bisecting frames are related by the same rotation about the y-axis $R_{\hat{\mathbf{y}}}(\theta)$, for both particles. In general by applying the same rotation to the parallel frames for both particles we can generate pairs of independent symmetrically-defined frames for any orientation of axes we like.

The relationship between the parallel frames of reference is a rotation of $\pm\pi$ about the axis $\hat{\mathbf{k}}$. This same rotation will also take one particle's vector into that of the other:
\begin{eqnarray}\label{eqn:asymrotparallel}
R(\mathbf{z}_a\rightarrow \mathbf{z}_b) & = R(\mathbf{v}_a\rightarrow \mathbf{v}_b) & = R_{ab} = R_{\mathbf{k}}(\pm\pi)\nonumber\\
R(\mathbf{z}_b\rightarrow \mathbf{z}_a) & = R(\mathbf{v}_b\rightarrow \mathbf{v}_a) & = R_{ba} = R_{ab}^{-1} = R_{\mathbf{k}}(\mp\pi)
\end{eqnarray}

A similar rotational relationship holds for the bisecting frames:
\begin{eqnarray}\label{eqn:asymrotbisect}
R(\mathbf{y}_a\rightarrow \mathbf{y}_b) & = R(\mathbf{v}_a\rightarrow \mathbf{v}_b) & = R_{ab} = R_{\mathbf{k}}(\pm\pi)\nonumber\\
R(\mathbf{y}_b\rightarrow \mathbf{y}_a) & = R(\mathbf{v}_b\rightarrow \mathbf{v}_a) & = R_{ba} = R_{ab}^{-1} = R_{\mathbf{k}}(\mp\pi)
\end{eqnarray}

In general, for any pair of symmetrically-defined independent frames the rotation which relates those frames is given by:
\begin{eqnarray}\label{eqn:asymrot}
R(\mathbf{v}_a\rightarrow \mathbf{v}_b) & = & R_{ab} = R_{\mathbf{k}}(\pm\pi)\nonumber\\
R(\mathbf{v}_b\rightarrow \mathbf{v}_a) & = & R_{ba} = R_{ab}^{-1} = R_{\mathbf{k}}(\mp\pi)
\end{eqnarray}

It doesn't matter whether we choose a clockwise rotation ($+\pi$) for $R_{ab}$ or an anti-clockwise rotation ($-\pi$): $R_{ba}$ will always be in the opposite direction.

Whatever pair of symmetrically-defined independent frames we choose, we may always select either frame as a common frame of reference as long as we explicitly account for the rotation of the independent frame of one particle into that of the other. But in doing so, because $R_{ab} \ne R_{ba}$, we break the symmetry. We can account for this asymmetry by explicitly including the rotation which takes one particle's vector into that of the other in the state description. Conventionally, however, this is not done. Hence, although such conventional  state vectors in this common frame will still be symmetric under permutations (if they are uniquely defined) the distinguishing labels necessitated by the selection of one particle's independent frame will introduce an exchange asymmetry.

This gives us the answer to the question we posed in section \ref{sec:asymcommon}: The asymmetry between $R_{ab}$ and $R_{ba}$ implies an asymmetry in any choice of common frame. However, if we always specify $R_a$ and $R_b$ with respect to the independent frames, such that $R_{ab}=R_{\mathbf{k}}(\pm\pi)$ then we have a means to handle this asymmetry in a common frame. By performing the appropriate rotations we can relate the exchange operation to permutation (under which the state vector is necessarily symmetric) compute the effect of exchanging the asymmetric distinguishing features and thereby obtain the exchange phase.

It is important to realize that the requirement $R_{ab}=R_{\mathbf{k}}(\pm\pi)$ for the relationship between the independent bisecting frames for a two-particle state vector applies even in the limit that $\hat{\mathbf{v}}_a$ and $\hat{\mathbf{v}}_b$ coincide. This is a crucial observation because, when $\hat{\mathbf{v}}_a=\hat{\mathbf{v}}_b$, it might be incautiously assumed that $R_{ab}$ was a null rotation. {\it It is because such an incautious assumption is made for the case of coincident spin quantization axes in the conventional construction, that the exchange phase of the ``spin-statistics'' theorem appears to be inexplicable as a result merely of permutation invariance}.

\section{Permutation Invariance In Momentum Space}
In this section we discuss uniqueness for single-particle state vectors of arbitrary momentum and spin. We then show how to use the prescription of the previous section to construct permutation-symmetric two-particle state vectors. We then show how the conventional construction is implicitly asymmetric and compute the exchange phase and the exclusion rule.

\subsection{Uniqueness And Spin Quantization In Momentum Space}
The purpose of this subsection is to review the definition of a momentum space state vector for particles of arbitrary spin and to illustrate and eliminate the ambiguities that might arise in order to ensure uniqueness of the state vector.

Conventionally, a state vector of arbitrary momentum $\mathbf{p}$ and spin component $m$ along an axis $\hat{\mathbf{n}}$ is defined by\cite{Wigner}:
\begin{equation}\label{eqn:wigdef}
|Q,\mathbf{p},s,m(\hat{\mathbf{n}})> = U(B(\mathbf{p}))\ |Q,\mathbf{0},s,m(\hat{\mathbf{n}})>
\end{equation}
where $|Q,\mathbf{0},s,m(\hat{\mathbf{n}})>$ is a rest frame eigenstate of $\mathbf{J}^2$ and component $J_{\hat{\mathbf{n}}}$ (in the direction $\hat{\mathbf{n}}$) with eigenvalues $s(s+1)$ and $m$, $U(B(\mathbf{p}))$ is the operator describing the boost (which in our case need not be a Lorentz boost, but could equally well be Galilean) which takes the momentum from $\mathbf{0}$ to $\mathbf{p}$:
\begin{equation}\label{eqn:boostdef}
U(B(\mathbf{p}))\ = U(R(\hat{\mathbf{z}}\rightarrow\hat{\mathbf{p}}))U(B(p\hat{\mathbf{z}}))U(R^{-1}(\hat{\mathbf{z}}\rightarrow\hat{\mathbf{p}}))
\end{equation}
and $Q$ represents all other intrinsic quantum numbers. We remind the reader, in passing, that the rotation $R(\hat{\mathbf{z}}\rightarrow\hat{\mathbf{p}})$ is defined to be that which takes the z-axis of the frame of reference from the direction of motion $\hat{\mathbf{p}}$ into the z-axis of the {\it final} frame of reference, therefore transforming the momentum from $p\hat{\mathbf{z}}$ to $\mathbf{p}$.

Two common choices of $\hat{\mathbf{n}}$ are:
\begin{enumerate}
\item The {\it canonical} basis ($\hat{\mathbf{n}} = \hat{\mathbf{z}}$) in which the spin quantization axis is the z-axis of the frame in which the momentum is measured.
\item The {\it helicity} basis ($\hat{\mathbf{n}} = \hat{\mathbf{p}}$) in which the quantization axis is parallel to the momentum.
\end{enumerate} 

Unfortunately, it isn't hard to see that both of these choices of state vector are potentially ambiguous up to an arbitrary rotation about their spin quantization axis, because none of the explicit state variables are changed by such a rotation. Even when we fix the other axes of the spin quantization frame, they are still ambiguous by a rotation by $2\pi$ about any axis. 

Under a rotation $R(\hat{\mathbf{n}}'\rightarrow \hat{\mathbf{n}})$ the rest frame vector transforms as (e.g. \cite{Rose}) 
\begin{eqnarray}
|Q,\mathbf{0},s,m(\hat{\mathbf{n}})> & = & U(R(\hat{\mathbf{n}}'\rightarrow \hat{\mathbf{n'}}))\ |Q,\mathbf{0},s,m(\hat{\mathbf{n}}')>\nonumber\\
& = & \sum_{m'}\ D^s_{m'm}(R(\hat{\mathbf{n}}'\rightarrow \hat{\mathbf{n}}))\ |Q,\mathbf{0},s,m'(\hat{\mathbf{n}}')>
\end{eqnarray}
and therefore, for general $\mathbf{p}$, state vectors with differing spin quantization frames are related by the rotation which relates those frames:
\begin{eqnarray}\label{eqn:rotatespinframe}
|Q,\mathbf{p},s,m(\hat{\mathbf{n}})>
& = & \sum_{m'}\ D^s_{m'm}(R(\hat{\mathbf{n}}'\rightarrow \hat{\mathbf{n}}))\ |Q,\mathbf{p},s,m'(\hat{\mathbf{n}}')>
\end{eqnarray}

Hence, a rotation of the spin quantization frame about the spin quantization axis $\hat{\mathbf{n}}$ will change the phase of the state vector, even though the spin quantization axis remains unchanged ($\hat{\mathbf{n}}'=\hat{\mathbf{n}}$). And if $R(\hat{\mathbf{n}}\rightarrow \hat{\mathbf{n}}')$ is a $2\pi$ rotation about any axis, the state vector changes phase by $(-1)^{2s}$, and so the specification of $\hat{\mathbf{n}}$ alone is not sufficient for a unique state vector. 

To remove the ambiguity we must uniquely specify the rotation which takes $\hat{\mathbf{z}}$ into $\hat{\mathbf{n}}$. We therefore define, instead of eqn. \ref{eqn:wigdef}
\begin{equation}\label{eqn:newwigdef}
|Q,\mathbf{p},s,m(R(\hat{\mathbf{z}}\rightarrow \hat{\mathbf{n}}))> = U(B(\mathbf{p}))\ |Q,\mathbf{0},s,m(R(\hat{\mathbf{z}}\rightarrow \hat{\mathbf{n}}))>
\end{equation}

In the canonical basis, $R(\hat{\mathbf{z}}\rightarrow \hat{\mathbf{n}})$ is limited (up to any additional rotation by $2\pi$ about any axis) to a rotation by $\alpha$ about the z-axis, where $\alpha$ has any value we choose. Normally, we would choose $\alpha = 0$, and define
\begin{eqnarray}
|Q,\mathbf{p},s,m>^C = |Q,\mathbf{p},s,m(N))>\label{eqn:canondef}
\end{eqnarray}
where $N$ is a null rotation, so that the spin quantization frame and the momentum frame of reference coincide. We note, in passing that the same coincidence of spin quantization frame and momentum reference frame is also true for
\begin{eqnarray}
|Q,\mathbf{p},s,m(R_{\hat{\mathbf{q}}}(2\pi))> = (-1)^{2s} |Q,\mathbf{p},s,m>^C
\end{eqnarray}
where $R_{\hat{\mathbf{q}}}(2\pi)$ is a rotation by $2\pi$ about any arbitrary axis $\hat{\mathbf{q}}$. However, by using the notation of eqn. \ref{eqn:newwigdef} and the definition of eqn. \ref{eqn:canondef} we avoid this phase ambiguity.

Similarly, in the helicity basis, we can define 
\begin{eqnarray}
|Q,\mathbf{p},s,\lambda>^H = |Q,\mathbf{p},s,\lambda(R(\hat{\mathbf{z}}\rightarrow \hat{\mathbf{p}}))>
\end{eqnarray}

The relationship between the canonical and helicity state vectors is then given by eqn. \ref{eqn:rotatespinframe}:
\begin{eqnarray}\label{eqn:heltocan}
|Q,\mathbf{p},s,m>^C
& = & \sum_{\lambda}\ D^s_{\lambda m}(R(\hat{\mathbf{p}}\rightarrow \hat{\mathbf{z}}))\ |Q,\mathbf{p},s,\lambda>^H
\end{eqnarray}

Under an arbitrary rotation R, which transforms the frame of reference in which the momentum is $\mathbf{p}$ into one in which the momentum is $\mathbf{p}'$, the general state vector defined in eqn. \ref{eqn:newwigdef} transforms according to:
\begin{eqnarray}
\lefteqn{U(R)|Q,\mathbf{p},s,m(R(\hat{\mathbf{z}}\rightarrow \hat{\mathbf{n}}))>}\nonumber\\ & = & U(B(\mathbf{p}'))U(R)|Q,\mathbf{0},m(R(\hat{\mathbf{z}}\rightarrow \hat{\mathbf{n}}))>\nonumber\\
& = & U(B(\mathbf{p}'))|Q,\mathbf{0},m(R(\hat{\mathbf{z}}\rightarrow \hat{\mathbf{n}}'))>\nonumber\\
& = & |Q,\mathbf{p}',s,m(R(\hat{\mathbf{z}}\rightarrow \hat{\mathbf{n}}'))>\nonumber\\
& = & \sum_{m'}\ D^s_{m'm}(R)\ |Q,\mathbf{p'},s,m'(R(\hat{\mathbf{z}}\rightarrow \hat{\mathbf{n}}))>
\end{eqnarray}
and 
\begin{eqnarray}
R(\hat{\mathbf{z}}\rightarrow \hat{\mathbf{n}}')=R(\hat{\mathbf{n}}\rightarrow \hat{\mathbf{n}}').R(\hat{\mathbf{z}}\rightarrow \hat{\mathbf{n}})=R(\hat{\mathbf{p}}\rightarrow \hat{\mathbf{p}}').R(\hat{\mathbf{z}}\rightarrow \hat{\mathbf{n}})=R.R(\hat{\mathbf{z}}\rightarrow \hat{\mathbf{n}})
\end{eqnarray}
is the rotation which takes the z-axis into the spin-quantization axis in the rotated system.

In the helicity basis, therefore:
\begin{eqnarray}
U(R) |Q,\mathbf{p},s,\lambda>^H = U(R)|Q,\mathbf{p},s,m(R(\hat{\mathbf{z}}\rightarrow \hat{\mathbf{p}}))>\nonumber\\
= |Q,\mathbf{p}',s,(R(\hat{\mathbf{z}}\rightarrow \hat{\mathbf{p}}'))> = |Q,\mathbf{p}',s,\lambda>^H
\end{eqnarray}
and the helicity is unchanged by the rotation.

Whereas, in the canonical basis:
\begin{eqnarray}
U(R)|Q,\mathbf{p},s,m>^C & = & U(R)|Q,\mathbf{p},s,m(N)>\nonumber\\
& = & \sum_{m'}\ D^s_{m'm}(R)\ |Q,\mathbf{p}',s,m'(N)>\nonumber\\
& = & \sum_{m'}\ D^s_{m'm}(R)\ |Q,\mathbf{p}',s,m'>^C
\end{eqnarray}
and we see that the third component of spin is transformed.

\subsection{Exchange Phase In Momentum Space}
Clearly, eqn. \ref{eqn:newwigdef} provides us with our desired uniqueness for single particle state vectors. However, we have seen from section \ref{sec:asymaccount} that a symmetric definition of two-particle state vectors requires us to specify independent frames of reference for each particle in a symmetric way. We also saw how to do this given two single-particle vectors in a way that enables a common z-axis. 

However, there are two additional problems which face us when it comes to doing this for the state vectors defined in the previous subsection. 

The first is that each particle has {\it two} vectors associated with it: its momentum $\mathbf{p}$ and its spin quantization axis $\mathbf{n}$. In general it would be difficult, or impossible, to symmetrically define independent frames of reference with respect to both momenta and spin quantization axes simultaneously. There are at least two special cases in which we {\it can} do this, however: the canonical basis and the helicity basis. In the canonical basis $\mathbf{n}_a=\mathbf{n}_b=\hat{\mathbf{z}}$, so it should be possible to choose a common z-axis bisecting $\mathbf{p}_a$ and $\mathbf{p}_b$. However, we saw in section \ref{sec:asymaccount} the importance of taking the general case ($\mathbf{n}_a\ne \mathbf{n}_b$) first. So we shall choose, for our first example, the helicity basis, and return to the canonical basis with this experience under our belt. Now since, in the helicity basis, $\mathbf{p}$ and $\mathbf{n}$ coincide, the independent frame for $\mathbf{p}$ is also the independent frame for $\mathbf{n}$. This enables us to define a two-particle helicity state vector with known symmetry, using independent frames. The relationship of eqn. \ref{eqn:heltocan} between the canonical basis and the helicity basis, then tells us that we can also define a two-particle canonical state vector  with known symmetry either by using eqn. \ref{eqn:heltocan} directly, or by repeating the method used for helicity states.

The second difficulty concerns  --- yet again --- ambiguous notation. For a single particle, the frame of reference is implicit in the rotation $R(\hat{\mathbf{z}}\rightarrow \hat{\mathbf{p}})$ which defines the angular co-ordinates of the momentum vector $\mathbf{p}$, or in the rotation $R(\hat{\mathbf{z}}\rightarrow \hat{\mathbf{n}})$, which defines the angular co-ordinates of the spin quantization axis $\mathbf{n}$, so it is not necessary to specify the frame of reference independently. When we come to a two-particle state vector, we must also specify how to relate the implicit frames of reference for each separate particle. We shall therefore use superscript labels for each particle to indicate unambiguously how its frame of reference has been chosen and, as and when it becomes necessary, we shall introduce notation to specify the relative orientation of these two frames. 

The independent bisecting frame, which we shall denote by $B_c$, for the {\it current} particle, is given by eqns. \ref{eqn:bisectingz} and \ref{eqn:currenty}:
\begin{eqnarray}
\hat{\mathbf{z}}_c & = & \hat{\mathbf{k}}\nonumber\\
\hat{\mathbf{y}}_c & = & \hat{\mathbf{p}}_c\times\hat{\mathbf{p}}_o
\end{eqnarray}
where $\mathbf{k}$ bisects $\mathbf{p}_a$ and $\mathbf{p}_b$. 

We then define the two-particle helicity state vector in the independent bisecting frames:
\begin{eqnarray}\label{eqn:xsymhel}
\lefteqn{|(Q_a,\mathbf{p}_a,s_a,\lambda_a)^{B_c};(Q_b,\mathbf{p}_b,s_b,\lambda_b)^{B_c}>^H}\nonumber\\
& = \alpha\ ( & |(Q_a,\mathbf{p}_a,s_a,\lambda_a)^{B_c}>^H|(Q_b,\mathbf{p}_b,s_b,\lambda_b)^{B_c}>^H\nonumber\\
& + & |(Q_b,\mathbf{p}_b,s_b,\lambda_b)^{B_c}>^H|(Q_a,\mathbf{p}_a,s_a,\lambda_a)^{B_c}>^H\ )\nonumber\\
& = \alpha\ ( & |(Q_a,\mathbf{p}_a,s_a,\lambda_a(R))^{B_c}>|(Q_b,\mathbf{p}_b,s_b,\lambda_b(R))^{B_c}>\nonumber\\
& + & |(Q_b,\mathbf{p}_b,s_b,\lambda_b(R))^{B_c}>|(Q_a,\mathbf{p}_a,s_a,\lambda_a(R))^{B_c}>\ )\nonumber\\
& = & |(Q_b,\mathbf{p}_b,s_b,\lambda_b)^{B_c};Q_a,\mathbf{p}_a,s_a,\lambda_a)^{B_c}>^H
\end{eqnarray}
where $R$ takes the bisecting frame $B_c$ of the current particle (z-axis is $\mathbf{k}$) into its helicity frame (z-axis is $\mathbf{p}_c$, measured in the frame of reference $B_c$) and, of course, is the same for both particles, because of the symmetry of the independent frames. Clearly, this state vector is permutation symmetric. Since the distinguishing superscripts are identical for both particles, then this state vector is also exchange symmetric. 

The rotation which transforms the frame $B_c$ into $B_o$ is given by $R_{co}=R(\hat{\mathbf{p}}_c\rightarrow \hat{\mathbf{p}}_o)=R_{\mathbf{k}}(\pm\pi)$, where $\mathbf{p}_c$ is measured in frame $B_c$ and $\mathbf{p}_o$ in frame $B_o$. The rotation $R(\hat{\mathbf{z}}\rightarrow \hat{\mathbf{p}}_o)$ which takes the z-axis of $B_o$ into $\hat{\mathbf{p}}_o$ is then given by $R(\hat{\mathbf{p}}_c\rightarrow \hat{\mathbf{p}}_o).R(\hat{\mathbf{z}}\rightarrow \hat{\mathbf{p}}_c)=R_{co}.R$. Hence:
\begin{eqnarray}
\lefteqn{U(R_{co})|(Q,\mathbf{p}_c,s,\lambda)^{B_c}>^H}\nonumber\\
& = & U(R_{co})|(Q,\mathbf{p}_c,s,\lambda(R))^{B_c}>\nonumber\\
& = & |(Q,\mathbf{p}_o,s,\lambda_b(R_{co}.R))^{B_o}>\nonumber\\
& = & |(Q,\mathbf{p}_o,s,\lambda_b(R_{\mathbf{k}}(\pm\pi).R))^{B_o}>\nonumber\\
& = & |(Q,\mathbf{p}_o,s,\lambda_b)^{B_o}>^H_{co,\pm}
\end{eqnarray}
and the suffices ``$co,\pm$'' in the last state vector indicate the value of the rotation chosen for $R_{co}$ which defines $B_o$ relative to $B_c$.

We can now define the helicity basis state vector for the common frame of reference given by the bisecting frame of particle $a$:
\begin{eqnarray}\label{eqn:xsym}
\lefteqn{|(Q_a,\mathbf{p}_a,s_a,\lambda_a)^{B_c};(Q_b,\mathbf{p}_b,s_b,\lambda_b)^{B_o}>^H_{co,\pm}}\nonumber\\
& = \alpha\ ( & |(Q_a,\mathbf{p}_a,s_a,\lambda_a)^{B_c}>^H|(Q_b,\mathbf{p}_b,s_b,\lambda_b)^{B_o}>^H_{\pm}\nonumber\\
& + & |(Q_b,\mathbf{p}_b,s_b,\lambda_b)^{B_o}>^H_{\pm}|(Q_a,\mathbf{p}_a,s_a,\lambda_a)^{B_c}>^H\ )\nonumber\\
& = \alpha\ ( & |(Q_a,\mathbf{p}_a,s_a,\lambda_a)^{B_c}>^H (U(R_{co})|(Q_b,\mathbf{p}_b',s_b,\lambda_b)^{B_c}>^H)\nonumber\\
& + & (U(R_{co})|(Q_b,\mathbf{p}_b',s_b,\lambda_b)^{B_c}>^H)|(Q_a,\mathbf{p}_a,s_a,\lambda_a)^{B_c}>^H\ )\nonumber\\
& = & |(Q_a,\mathbf{p}_a,s_a,\lambda_a(R))^{B_c};(Q_b,\mathbf{p}_b,s_b,\lambda_b(R_{co}.R))^{B_o}>\nonumber\\
& = & |(Q_a,\mathbf{p}_a,s_a,\lambda_a(R))^{B_c};(Q_b,\mathbf{p}_b,s_b,\lambda_b(R_{\mathbf{k}}(\pm\pi).R))^{B_o}>
\end{eqnarray}
which is again symmetric under particle permutation. However, particle exchange is more complex, since the transformation $B_c\leftrightarrow B_o$ for each particle will result in a change in the common frame of reference from the bisecting frame of one particle to that of the other, and therefore will, in general, involve a change in each particle's momentum rather than just a change in phase. Furthermore, since $R_{oc} = R_{co}^{-1}$, each particle undergoes a different rotation:
\begin{eqnarray}
\lefteqn{|(Q_b,\mathbf{p}_b,s_b,\lambda_b)^{B_c};(Q_a,\mathbf{p}_a,s_a,\lambda_a)^{B_o}>^H_{co,\pm}}\nonumber\\
& = & |(Q_b,\mathbf{p}_b,s_b,\lambda_b(R))^{B_c};(Q_a,\mathbf{p}_a,s_a,\lambda_a(R_{co}.R))^{B_o}>\nonumber\\
& = \alpha\ ( & (U(R_{co})|(Q_a,\mathbf{p}_a',s_a,\lambda_a)^{B_c}>^H)
(U(R_{oc})|(Q_b,\mathbf{p}_b',s_b,\lambda_b)^{B_o}>^H_{co,\pm})\nonumber\\
& + & (U(R_{oc})|(Q_b,\mathbf{p}_b',s_b,\lambda_b)^{B_o}>^H_{co,\pm}) (U(R_{co})|(Q_a,\mathbf{p}_a',s_a,\lambda_a)^{B_c}>^H)\ )\nonumber\\
\end{eqnarray}
and since $R_{oc}=R_{co}.R_{oc}.R_{oc} = R_{co}.R_{\mathbf{k}}(\pm 2\pi)$, we find:
\begin{eqnarray}\label{eqn:xasym}
\lefteqn{|(Q_b,\mathbf{p}_b,s_b,\lambda_b)^{B_c};(Q_a,\mathbf{p}_a,s_a,\lambda_a)^{B_o}>^H_{co,\pm}}\nonumber\\
& = & (-1)^{2s_b}U(R_{co})
|(Q_a,\mathbf{p}_a',s_a,\lambda_a)^{B_c};(Q_b,\mathbf{p}_b',s_b,\lambda_b)^{B_o}>^H_{co,\pm}\nonumber\\
& = & (-1)^{2s_a}U(R_{oc})
|(Q_a,\mathbf{p}_a',s_a,\lambda_a)^{B_c};(Q_b,\mathbf{p}_b',s_b,\lambda_b)^{B_o}>^H_{co,\pm}
\end{eqnarray}

In any other frame of reference $F$, given by the rotation $R^F_c=R(\hat{\mathbf{p}}_c\rightarrow \hat{\mathbf{p}})$, which transforms the frame $B_c$ into $F$ (where $\mathbf{p}_c$ measured in $B_c$ becomes $\mathbf{p}$ measured in $F$ or where $\mathbf{p}_o$ measured in $B_o$ becomes $\mathbf{p}$ measured in $F$), we define:
\begin{eqnarray}
\lefteqn{|(Q_a,\mathbf{p}_a,s_a,\lambda_a)^{1};(Q_b,\mathbf{p}_b,s_b,\lambda_b)^{2}>^H_\pm}\nonumber\\
& = & U(R^F_a)
|(Q_a,\mathbf{p}_a',s_a,\lambda_a)^{B_c};(Q_b,\mathbf{p}_b',s_b,\lambda_b)^{B_o}>^H_{co,\pm}\nonumber\\
& = & |(Q_b,\mathbf{p}_b,s_b,\lambda_b)^{2};(Q_a,\mathbf{p}_a,s_a,\lambda_a)^{1}>^H_\pm
\end{eqnarray}
where the superscript label ``1'' indicates which particle's frame of reference was its {\it current} frame when we applied the rotation $R^F_c$ and the label ``2'' indicates which particle's frame of reference was its {\it other} frame, related to its own {\it current} frame by the rotation $R_{co}$ --- the value of which is specified, once more, by the suffix``$\pm$''. This state vector, is of course, once again, permutation symmetric. But under particle exchange we have:
\begin{eqnarray}
\lefteqn{|(Q_b,\mathbf{p}_b,s_b,\lambda_b)^{1};(Q_a,\mathbf{p}_a,s_a,\lambda_a)^{2}>^H_\pm}\nonumber\\
& = & U(R^F_b)
|(Q_b,\mathbf{p}_b'',s_b,\lambda_b)^{B_c};(Q_a,\mathbf{p}_a'',s_a,\lambda_a)^{B_o}>^H_{co,\pm}\nonumber\\
& = & (-1)^{2s_a}U(R^F_b.R_{oc})
|(Q_a,\mathbf{p}_a',s_a,\lambda_a)^{B_c};(Q_b,\mathbf{p}_b',s_b,\lambda_b)^{B_o}>^H_{co,\pm}\nonumber\\
& = & (-1)^{2s_b}U(R^F_b.R_{co})
|(Q_a,\mathbf{p}_a',s_a,\lambda_a)^{B_c};(Q_b,\mathbf{p}_b',s_b,\lambda_b)^{B_o}>^H_{co,\pm}
\end{eqnarray}
where the last steps follow from eqn. \ref{eqn:xasym}. Now, $R^F_b= R^F_a.R_{ba}$ ($R_{ba} = (R^F_a)^{-1}.R^F_b$) and $R_{ba}=R_{\mathbf{k}}(\pm 2\pi)$. Hence either $R_{ba}=R_{co}$ or $R_{ba}=R_{oc}$, since it rotates the bisecting frame of $b$ into that of $a$. Choosing $R_{ba}=R_{co}$, we find:
\begin{eqnarray}\label{eqn:xasymFa}
\lefteqn{|(Q_b,\mathbf{p}_b,s_b,\lambda_b)^{1};(Q_a,\mathbf{p}_a,s_a,\lambda_a)^{2}>^H_\pm}\nonumber\\
& = & (-1)^{2s_a}U(R^F_a)
|(Q_a,\mathbf{p}_a',s_a,\lambda_a)^{B_c};(Q_b,\mathbf{p}_b',s_b,\lambda_b)^{B_o}>^H_{co,\pm}\nonumber\\
& = & (-1)^{2s_a}
|(Q_a,\mathbf{p}_a,s_a,\lambda_a)^{1};(Q_b,\mathbf{p}_b,s_b,\lambda_b)^{2}>^H_\pm
\end{eqnarray}
But if we choose $R_{ba}=R_{oc}$, then
\begin{eqnarray}\label{eqn:xasymFb}
\lefteqn{|(Q_b,\mathbf{p}_b,s_b,\lambda_b)^{1};(Q_a,\mathbf{p}_a,s_a,\lambda_a)^{2}>^H_\pm}\nonumber\\
& = & (-1)^{2s_b}
|(Q_a,\mathbf{p}_a,s_a,\lambda_a)^{1};(Q_b,\mathbf{p}_b,s_b,\lambda_b)^{2}>^H_\pm
\end{eqnarray}
and now, the significance of the superscript labels ``1'' and ``2'', which specifies that $B_2$ was rotated by $R_{co}$ to $B_1$ before rotating $B_1$ to $F$, is that the state vector in a fixed common frame has an exchange phase given by a $2\pi$ rotation on the frame of reference of one of the particles --- which particle depending on how we define $R^F_{ab}$ relative to $R_{oc}$. Clearly, in the case that both particles are bosons or both fermions, there is no difference between the exchange phase in eqns. \ref{eqn:xasymFa} and \ref{eqn:xasymFb}. In the case of identical particles, $s_a=s_b=s$, the exchange phase is always $(-1)^{2s}$ in a common frame of reference whatever choice we make for $R_{ab}$. (For the sake of tying up loose ends, we would mention that the difference in the exchange phases between eqn. \ref{eqn:xasymFa} and eqn. \ref{eqn:xasymFb} reflects the overall ambiguity in specifying the frame of reference $F$. Applying an additional $2\pi$ rotation to the common frame of reference (or, alternatively, rotating the spin quantization frames) will result in an additional phase factor $(-1)^{2s_a+2s_b}$.)

Now let us turn our attention to the canonical basis. We follow an analogous procedure to that used for the helicity basis. Once again, using the independent bisecting frames defined by eqns. \ref{eqn:bisectingz} and \ref{eqn:currenty}, we define:
\begin{eqnarray}\label{eqn:xsymcan}
\lefteqn{|(Q_a,\mathbf{p}_a,s_a,m_a)^{B_c};(Q_b,\mathbf{p}_b,s_b,m_b)^{B_c}>^C}\nonumber\\
& = \alpha\ ( & |(Q_a,\mathbf{p}_a,s_a,m_a)^{B_c}>^C|(Q_b,\mathbf{p}_b,s_b,m_b)^{B_c}>^C\nonumber\\
& + & |(Q_b,\mathbf{p}_b,s_b,m_b)^{B_c}>^C|(Q_a,\mathbf{p}_a,s_a,m_a)^{B_c}>^C\ )\nonumber\\
& = \alpha\ ( &  |(Q_a,\mathbf{p}_a,s_a,m_a(N))^{B_c}>|(Q_b,\mathbf{p}_b,s_b,m_b(N))^{B_c}>\nonumber\\
& + & |(Q_b,\mathbf{p}_b,s_b,m_b(N))^{B_c}>|(Q_a,\mathbf{p}_a,s_a,m_a(N))^{B_c}>\ )\nonumber\\
& = & |(Q_b,\mathbf{p}_b,s_b,m_b)^{B_c};(Q_a,\mathbf{p}_a,s_a,m_a)^{B_c}>^C
\end{eqnarray}
where $N$ is a null rotation. As before, this state vector is permutation and exchange symmetric. Note that since the bisecting frames do not coincide, neither do the spin quantization frames.

Now, in any frame $F$, applying a rotation $R$ has the effect:
\begin{equation}
U(R)|(Q,\mathbf{p},s,m(R^F))^F> =  |(Q,\mathbf{p}',s,m(R.R^F))^{F'}>
\end{equation}

Applying the rotation $R_{co}$ to canonical frame $B_c$, we get:
\begin{eqnarray}\label{eqn:rotcanonc2o}
\lefteqn{U(R_{co})|(Q,\mathbf{p}_c,s,m)^{B_c}>^C}\nonumber\\
& = & U(R_{co})|(Q,\mathbf{p}_c,s,m(N)))^{B_c}>\nonumber\\
& = & |(Q,\mathbf{p}_o,s,m(R_{co}))^{B_o}>
\end{eqnarray}

Similarly, applying $R_{oc}$ to canonical frame $B_o$:
\begin{eqnarray}\label{eqn:rotcanono2c}
\lefteqn{U(R_{oc})|(Q,\mathbf{p}_o,s,m)^{B_o}>^C_{co,\pm}}\nonumber\\
& = & U(R_{oc})|(Q,\mathbf{p}_o,s,m(N))^{B_o}>_{co,\pm}\nonumber\\
& = & |(Q,\mathbf{p}_c,s,m(R_{oc}))^{B_c}>
\end{eqnarray}
and, again, the suffices ``$co,\pm$'' refer to the value of $R_{co}$ chosen to relate $B_o$ to $B_c$ for a state vector defined in $B_o$.

Applying the rotation $R_{co}$ to both sides of this eqn. \ref{eqn:rotcanono2c} gives:
\begin{eqnarray}\label{eqn:roto2c}
|(Q,\mathbf{p}_o,s,m)^{B_o}>^C_{co,\pm}
& = & U(R_{co})|(Q,\mathbf{p}_c,s,m(R_{oc}))^{B_c}>
\end{eqnarray}

Hence, we find, in the common frame $B_a$ which is also the common spin quantization frame:
\begin{eqnarray}\label{eqn:xsymcana}
\lefteqn{|(Q_a,\mathbf{p}_a,s_a,m_a)^{B_c};(Q_b,\mathbf{p}_b,s_b,m_b)^{B_o}>^C_{co,\pm}}\nonumber\\
& = &  |(Q_a,\mathbf{p}_a,s_a,m_a(N))^{B_c};(Q_b,\mathbf{p}_b,s_b,m_b(N))^{B_o}>_{co,\pm}\nonumber\\
& = & \alpha\ ( |(Q_a,\mathbf{p}_a,s_a,m_a(N))^{B_c}>
(U(R_{co}) |(Q_b,\mathbf{p}_b',s_b,m_b(R_{oc}))^{B_c}>)\nonumber\\
& & + (U(R_{co}) |(Q_b,\mathbf{p}_b',s_b,m_b(R_{oc})^{B_c}>)
|(Q_a,\mathbf{p}_a,s_a,m_a(N))^{B_c}>\ )
\end{eqnarray}
where we have used eqn. \ref{eqn:roto2c} to relate each particle's state to its bisecting frame. 

As usual, this is permutation symmetric. Under particle exchange:
\begin{eqnarray}\label{eqn:xsymcanb}
\lefteqn{|(Q_b,\mathbf{p}_b,s_b,m_b)^{B_c};(Q_a,\mathbf{p}_a,s_a,m_a)^{B_o}>^C_{co,\pm}}\nonumber\\
& = & |(Q_b,\mathbf{p}_b,s_b,m_b(N))^{B_c};(Q_a,\mathbf{p}_a,s_a,m_a(N))^{B_o}>_{co,\pm}\nonumber\\
& = & \alpha\ (  (U(R_{oc})|(Q_b,\mathbf{p}_b',s_b,m_b(R_{co}))^{B_o}>_{co,\pm})(U(R_{co})|(Q_a,\mathbf{p}_a',s_a,m_a(R_{oc}))^{B_c}>)\nonumber\\
& & +(U(R_{co})|(Q_a,\mathbf{p}_a',s_a,m_a(R_{oc}))^{B_c}>)(U(R_{oc})|(Q_b,\mathbf{p}_b',s_b,m_b(R_{co}))^{B_o}>_{co,\pm})\ )\nonumber\\
& = & (-1)^{\pm(m_a+m_b)} U(R_{oc})
|(Q_a,\mathbf{p}_a',s_a,m_a)^{B_c};(Q_b,\mathbf{p}_b',s_b,m_b)^{B_o}>^C_{co,\pm}\nonumber\\
& = & (-1)^{\mp(m_a+m_b)} U(R_{co})
|(Q_a,\mathbf{p}_a',s_a,m_a)^{B_c};(Q_b,\mathbf{p}_b',s_b,m_b)^{B_o}>^C_{co,\pm}
\end{eqnarray}
and we see that exchange is an overall rotation of the frame of reference from $B_a$ to $B_b$ followed by the same rotation on the spin quantization frames, the latter resulting in the phase factor shown.

In a general canonical frame $F$, related to the independent bisecting frame $B_c$ by a rotation $R^F_c$, we have:
\begin{eqnarray}
|(Q,\mathbf{p},s,m)^F>^C = |(Q,\mathbf{p},s,m(N))^F> = U(R^F_c)
|(Q,\mathbf{p}',s,m((R^F_c)^{-1}))^{B_c}>
\end{eqnarray}

Similarly,
\begin{eqnarray}
|(Q,\mathbf{p},s,m)^F>^C = U(R^F_o)|(Q,\mathbf{p}',s,m((R^F_o)^{-1}))^{B_o}>_{co,\pm}
\end{eqnarray}

In the common frame $F$ which is also the common spin quantization frame, we therefore define:
\begin{eqnarray}\label{eqn:xasymFacanon}
\lefteqn{|(Q_a,\mathbf{p}_a,s_a,m_a)^1;(Q_b,\mathbf{p}_b,s_b,m_b)^2>^C_\pm}\nonumber\\
& = & U(R^F_a) \nonumber\\
& & |(Q_a,\mathbf{p}_a,s_a,m_a((R^F_a)^{-1}))^{B_c};(Q_b,\mathbf{p}_b,s_b,m_b((R^F_a)^{-1}))^{B_o}>_{co,\pm}
\end{eqnarray}
which, as usual, is permutation symmetric. As before, we exchange ``1'' and ``2'' to find, in the same common canonical frame $F$:
\begin{eqnarray}\label{eqn:xasymFbcanon}
\lefteqn{|(Q_b,\mathbf{p}_b,s_b,m_b)^{1};(Q_a,\mathbf{p}_a,s_a,m_a)^{2}>^C_\pm}\nonumber\\
& = & U(R^F_b)
|(Q_b,\mathbf{p}_b,s_b,m_b((R^F_b)^{-1}))^{B_c};(Q_a,\mathbf{p}_a,s_a,m_a((R^F_b)^{-1}))^{B_o}>_{co,\pm}\nonumber\\
& = & \alpha\ ( 
(U(R^F_b.R_{co})|(Q_a,\mathbf{p}_a,s_a,m_a((R^F_b.R_{co})^{-1}))^{B_c}>)\nonumber\\
& . & (U(R^F_b.R_{oc})|(Q_b,\mathbf{p}_b,s_b,m_b((R^F_b.R_{oc})^{-1}))^{B_o}>_{co,\pm})\nonumber\\
& + & 
(U(R^F_b.R_{oc})|(Q_b,\mathbf{p}_b,s_b,m_b((R^F_b.R_{oc})^{-1}))^{B_o}>_{co,\pm})\nonumber\\
& . & (U(R^F_b.R_{co})|(Q_a,\mathbf{p}_a,s_a,m_a((R^F_b.R_{co})^{-1}))^{B_c}>)
\ )\nonumber\\
& = & \alpha\ ( 
(U(R^F_a.R_{ba}.R_{co})|(Q_a,\mathbf{p}_a,s_a,m_a((R^F_a.R_{ba}.R_{co})^{-1}))^{B_c}>)\nonumber\\
& . & (U(R^F_a.R_{ba}.R_{oc})|(Q_b,\mathbf{p}_b,s_b,m_b((R_a.R^F_{ba}.R_{oc})^{-1}))^{B_o}>_{co,\pm})\nonumber\\
& + & 
(U(R^F_a.R_{ba}.R_{oc})|(Q_b,\mathbf{p}_b,s_b,m_b((R^F_a.R_{ba}.R_{oc})^{-1}))^{B_o}>_{co,\pm})\nonumber\\
& . & (U(R^F_a.R_{ba}.R_{co})|(Q_a,\mathbf{p}_a,s_a,m_a((R^F_a.R_{ba}.R_{co})^{-1}))^{B_c}>)
\ )
\end{eqnarray}

As before, if we choose $R_{ba} = R_{co}$ then:
\begin{eqnarray}\label{eqn:xasymcanon1}
\lefteqn{|(Q_b,\mathbf{p}_b,s_b,m_b)^{1};(Q_a,\mathbf{p}_a,s_a,m_a)^{2}>^C_\pm}\nonumber\\
& = & \alpha\ (
(U(R^F_a.R_{\mathbf{k}}(\pm 2\pi))|(Q_a,\mathbf{p}_a,s_a,m_a((R^F_a.R_{\mathbf{k}}(\pm 2\pi))^{-1}))^{B_c}>)\nonumber\\
& . & (U(R^F_a)|(Q_b,\mathbf{p}_b,s_b,m_b((R^F_a)^{-1}))^{B_o}>_{co,\pm})\nonumber\\
& + & 
(U(R^F_a)|(Q_b,\mathbf{p}_b,s_b,m_b((R^F_a)^{-1}))^{B_o}>_{co,\pm})\nonumber\\
& . &(U(R^F_a.R_{\mathbf{k}}(\pm 2\pi))|(Q_a,\mathbf{p}_a,s_a,m_a((R^F_a.R_{\mathbf{k}}(\pm 2\pi))^{-1}))^{B_c}>)
\ )\nonumber\\
& = & (-1)^{2s_a} U(R^F_a)\nonumber\\
& & |(Q_a,\mathbf{p}_a,s_a,m_a((R^F_a)^{-1}))^{B_c};(Q_b,\mathbf{p}_b,s_b,m_b((R^F_a)^{-1}))^{B_o}>_{co,\pm}\nonumber\\
& = & (-1)^{2s_a} |(Q_a,\mathbf{p}_a,s_a,m_a)^{1};(Q_b,\mathbf{p}_b,s_b,m_b)^{2}>^C_\pm
\end{eqnarray}
and, if we choose $R_{ba} = R_{oc}$ then:
\begin{eqnarray}\label{eqn:xasymcanon2}
\lefteqn{|(Q_b,\mathbf{p}_b,s_b,m_b)^{1};(Q_a,\mathbf{p}_a,s_a,m_a)^{2}>^C_\pm}\nonumber\\
& = & (-1)^{2s_b}|(Q_a,\mathbf{p}_a,s_a,m_a)^{1};(Q_b,\mathbf{p}_b,s_b,m_b)^{2}>^C_\pm
\end{eqnarray}

Now it is important to remember that this exchange asymmetry arose because we used a fixed value of $R_{co}$, whichever particle was current. As we have seen before, this is not consistent with a fixed asymmetry of $R_{ab}\ne R_{ba}$, where $R_{ab}$ takes the bisecting frame of $a$ into that of $b$, unless we apply a rotation by $\pm 2\pi$ to the frame of reference of one of the particles and this rotation is clearly the origin of the non-vanishing exchange phase. 

However, it is quite possible to define state vectors that are {\it always} exchange symmetric in any frame of reference (or pair of independent frames of reference) and any choice of spin quantization axes by using a fixed value of $R_{ab}$ instead of a fixed value of $R_{co}$ (and keeping it unchanged under exchange).

For instance, instead of eqn. \ref{eqn:roto2c}, we define:
\begin{eqnarray}
|(Q_b,\mathbf{p}_b,s_b,m_b(R))^{B_a}>_{ba,\pm}
& = & U(R_{ba})|(Q_b,\mathbf{p}_b',s_b,m_b(R_{ab}.R)))^{B_b}>\nonumber\\
\end{eqnarray}
where, this time, the suffices ``$ba,\pm$'' imply a specific value of $R_{ba}$ has been chosen to define the state vector for particle $b$ in the bisecting frame of reference of particle $a$.

Then, in any common frame of reference $F$, we define the state vector:
\begin{eqnarray}
\lefteqn{|(Q_a,\mathbf{p}_a,s_a,m_a(R_a))^F;(Q_b,\mathbf{p}_b,s_b,m_b(R_b))^F>_{ba,\pm}}\nonumber\\
& = & U(R^F_a)|(Q_a,\mathbf{p}_a,s_a,m_a((R^F_a)^{-1}.R_a))^{B_a};
(Q_b,\mathbf{p}_b,s_b,m_b((R^F_a)^{-1}).R_b)^{B_a}>_{ba,\pm}\nonumber\\
& = & \alpha \ U(R^F_a) (\nonumber\\
& & |(Q_a,\mathbf{p}_a,s_a,m_a((R^F_a)^{-1}.R_a))^{B_a}>\nonumber\\
& . & |(Q_b,\mathbf{p}_b,s_b,m_b((R^F_a)^{-1}).R_b)^{B_a}>_{ba,\pm}\nonumber\\
& + & \ |(Q_b,\mathbf{p}_b,s_b,m_b((R^F_a)^{-1}).R_b)^{B_a}>_{ba,\pm}\nonumber\\
& . & |(Q_a,\mathbf{p}_a,s_a,m_a((R^F_a)^{-1}).R_a)^{B_a}>\ )\nonumber\\
& = & \alpha \ U(R^F_a) (\nonumber\\
& & |(Q_a,\mathbf{p}_a,s_a,m_a((R^F_a)^{-1}.R_a))^{B_a}>\nonumber\\
& . & (U(R_{ba})|(Q_b,\mathbf{p}_b,s_b,m_b((R^F_b)^{-1}).R_b)^{B_b}>)\nonumber\\
& + & (U(R_{ba})|(Q_b,\mathbf{p}_b,s_b,m_b((R^F_b)^{-1}).R_b)^{B_b}>)\nonumber\\
& . & |(Q_a,\mathbf{p}_a,s_a,m_a((R^F_a)^{-1}).R_a)^{B_a}>\ )
\end{eqnarray}
where $R_a$ and $R_b$ are the rotations which take common frame $F$ into the spin quantization frames of particle $a$ and $b$ respectively and $R^F_b = R^F_a.R_{ba}$. Since the (superscript) frame of reference is the same for both particles, and the fixed rotation $R_{ba}$ is unchanged by the exchange, this state vector is both permutation and exchange symmetric.

In particular, in the canonical basis, where $R_a=R_b=N$:
\begin{eqnarray}\label{eqn:xsymcanon}
\lefteqn{|(Q_a,\mathbf{p}_a,s_a,m_a)^F;(Q_b,\mathbf{p}_b,s_b,m_b)^F>^C_{ba,\pm}}\nonumber\\
& = & |(Q_a,\mathbf{p}_a,s_a,m_a(N))^F;(Q_b,\mathbf{p}_b,s_b,m_b(N))^F>_{ba,\pm}\nonumber\\
& = & U(R^F_a)\nonumber\\
& & \ |(Q_a,\mathbf{p}_a,s_a,m_a((R^F_a)^{-1}))^{B_a};
(Q_b,\mathbf{p}_b,s_b,m_b((R^F_a)^{-1}))^{B_a}>_{ba,\pm}\nonumber\\
& = & |(Q_b,\mathbf{p}_b,s_b,m_b)^F;(Q_a,\mathbf{p}_a,s_a,m_a)^F>^C_{ba,\pm}
\end{eqnarray}
is also permutation and exchange symmetric by virtue of the condition that $R_{ba}$ is fixed.

In the conventional construction, no explicit reference is made to the rotation $R_{ba}$ or to any particular choice for $R_{ba}$ since no care is taken to specify the rotation which relates $B_b$ to $B_a$. In the absence of an explicit choice of $R_{ba}$, uniqueness requires a fixed value of $R_{21}$ (relating the frame $B_2$ to the frame $B_1$ which, unlike $R_{ba}$ does not change under $b\leftrightarrow a$). Clearly this is satisfied by eqn. \ref{eqn:xasymFacanon}, where $R_{21}=R_{co}$. Therefore, the conventional construction corresponds to the state vectors of eqn. \ref{eqn:xasymcanon2} rather than eqn. \ref{eqn:xsymcanon}. However, it doesn't matter which we use, since we know how to unambiguously relate one definition to the other by specifying the values of $R_{ba}$ and $R_{co}$.

Specifically, the relationship between the order-dependent construction and the order-independent construction is:
\begin{eqnarray}
\lefteqn{|(Q_a,\mathbf{p}_a,s_a,m_a)^1;(Q_b,\mathbf{p}_b,s_b,m_b)^2>^C_\pm}\nonumber\\
& = & |(Q_a,\mathbf{p}_a,s_a,m_a)^F;(Q_b,\mathbf{p}_b,s_b,m_b)^F>^C_{ba,\pm}
\end{eqnarray}

\subsection{The Exclusion Rules}
\label{sec:xrules}

Having defined both exchange symmetric and exchange asymmetric state vectors for arbitrary frames of reference, it remains to determine the exclusion rules.

To do this, we need identical particle states of definite total spin $S$. Clearly, the quantization of the total spin requires a {\it single} unique choice of quantization frame.  To combine angular momentum, therefore, we need a common spin quantization frame for both particles, or, equivalently, must know how to relate the individual spin quantization frames to the total spin quantization frame.

For simplicity we shall use state vectors that employ the same spin quantization frame for both particles. We start with the order-dependent state vectors of eqn. \ref{eqn:xasymFacanon} which obey (from eqns. \ref{eqn:xasymcanon1} and \ref{eqn:xasymcanon2}):
\begin{eqnarray}
\lefteqn{|(Q,\mathbf{p}_b,s,m_b)^{1};(Q,\mathbf{p}_a,s,m_a)^{2}>^C_\pm}\nonumber\\
& = & (-1)^{2s}
|(Q,\mathbf{p}_a,s,m_a)^{1};(Q,\mathbf{p}_b,s,m_b)^{2}>^C_\pm
\end{eqnarray}

The state vector for states of total spin S and third component M are then given by:
\begin{eqnarray}
\lefteqn{|(S,M):(Q,\mathbf{p}_b,s)^{1};(Q,\mathbf{p}_a,s)^{2}>^C_\pm}\nonumber\\
& = & \sum_{m_a,m_b} C^{ssS}_{m_am_bM}|(Q,\mathbf{p}_a,s,m_a)^{1};(Q,\mathbf{p}_b,s,m_b)^{2}>^C_\pm
\end{eqnarray}

Then from the symmetry of the Clebsch-Gordon coefficients\cite{Rose}:
\begin{equation}
C^{ssS}_{m_am_bM} = (-1)^{S-2s} C^{ssS}_{m_bm_aM}
\end{equation}
we determine that, in the case $\mathbf{p}_a=\mathbf{p}_b = \mathbf{p}$:
\begin{eqnarray}
\lefteqn{|(S,M):(Q,\mathbf{p},s)^{1};(Q,\mathbf{p},s)^{2}>^C_\pm}\nonumber\\
& = & (-1)^S
|(S,M):(Q,\mathbf{p},s)^{1};(Q,\mathbf{p},s)^{2}>^C_\pm
\end{eqnarray}
from which we see that states of odd $S$ have vanishing state vectors and are therefore excluded.

\section{Acknowledgment}

The author would like to thank Prof. M. Berry for very useful, challenging and encouraging conversations which have helped him greatly with the presentation of this article.


\begin{thebibliography}{99}
\bibitem{Duck&Sud}Ian Duck \& E. C. G. Sudershan, \textsl{Toward an understanding of the spin-statistics theorem}, Am. J. Phys. {\bf 66} (4), April 1998
\bibitem{Berry&Rob} M. V. Berry \& J. M. Robbins, \textsl{Indistinguishability for Quantum Particles: Spin, Statistics and the Geometric Phase}, Proc. R. Soc. London Ser. A {\bf 453}, 1771-1790 (1997)
\bibitem{Rose} M. E. Rose \textsl{Elementary Theory Of Angular Momentum}, John Wiley \& Sons (1957) and Dover Publications Inc. (1995)
\bibitem{Wigner} E. P. Wigner, Rev. Modern Phys. {\bf 29}, 255 (1957)
\end{thebibliography}
\end{document}